\documentclass[10pt,letterpaper]{article}

\usepackage{ccn}
\usepackage{pslatex}
\usepackage{apacite}
\usepackage{graphicx}
\graphicspath{ {images/} }
\usepackage{amsmath}
\usepackage{setspace}
\usepackage{color,soul}

\title{The functional role of cue-driven feature-based feedback in object recognition }

\author{{\large \bf Sushrut Thorat \qquad Marcel van Gerven \qquad Marius Peelen} \\
sushrut.thorat94@gmail.com, m.vangerven@donders.ru.nl, m.peelen@donders.ru.nl\\
  Donders Institute for Brain, Cognition and Behaviour, Radboud University}

\begin{document}

\maketitle

\section{Abstract}
{
\bf
Visual object recognition is not a trivial task, especially when the objects are degraded or surrounded by clutter or presented briefly. External cues (such as verbal cues or visual context) can boost recognition performance in such conditions. In this work, we build an artificial neural network to model the interaction between the object processing stream (OPS) and the cue. We study the effects of varying neural and representational capacities of the OPS on the performance boost provided by cue-driven feature-based feedback in the OPS. We observe that the feedback provides performance boosts only if the category-specific features about the objects cannot be fully represented in the OPS. This representational limit is more dependent on task demands than neural capacity. We also observe that the feedback scheme trained to maximise recognition performance boost is not the same as tuning-based feedback, and actually performs better than tuning-based feedback.}
\begin{quote}
\small
\textbf{Keywords:} 
feature-based feedback; vision; neural networks
\end{quote}

\section{Introduction}

Visual object recognition is a non-trivial task, especially when the objects are degraded, surrounded by clutter or presented briefly. The introduction of external cues (such as verbal cues or visual context) can constrain the space of the possible object features and/or categories and improve recognition performance~\cite{carrasco2011visual,bar2004visual}. 

External cues can interact with the object processing stream in two ways. They can either modulate the information transformations in the object processing stream (through feedback) and/or get combined with the object evidence present at the end of the stream, to improve the overall decision about the category of the object. Intuitively, the interaction involving feedback would help with object recognition especially when the feature information required to recognise the object cannot be extracted by the object processing stream. This can happen either due to a capacity limit or due to a lack of information present in the input.

Such a capacity limit can arise due to two reasons. One, due to a limit on the number of neurons available in the object processing stream, which would reduce the object information that can be extracted from the image. We term this the \textit{neural} capacity limit. Two, due to the limits imposed by the task for which the stream is trained. For example, if the stream is trained to represent one object, it will not perform well if two objects are presented in the same image unless feature selection is employed in the early stages of the network. We term this the \textit{representational} capacity limit. In this work, we aim to understand how the feedback-driven performance gain due to the cue depends on the capacity limits of the visual processing stream. 

Cue-driven feedback can affect the object processing stream in a feature-specific and/or a location-specific manner. These interactions also account for feature-based and spatial attention~\cite{carrasco2011visual}. In this work, we focus on the feature-based feedback interaction.

We developed an artificial neural network (ANN) to model the object processing stream, probing the stream output for an object's presence in the image, and the feedback-based interaction between an external cue and the stream. The parameters of the ANN let us manipulate the neural and representational capacities of the object processing stream. We train the ANN to maximise the difference (termed as the recognition performance) between correct (true positives) and incorrect (false positives) identification of the objects in the image. The external cue and the probe contain category-level information. For example, the external cue could correspond to 'Look for a Shoe' and the probe could correspond to 'Was there a Shoe?'. Then the category-level information would be `Shoe'. We show that the external cue substantially boosts recognition performance when the object processing stream cannot represent (low representational capacity) the information required for categorising the target object (given by the probe). We then comment on the nature of the feedback that maximises these performance boosts.

\section{Methods}

\subsection{Stimuli}
\label{sec:stim}

We use the dataset Fashion-MNIST~\cite{xiao2017fashion}, which contains $28\times 28$ images of $70,000$ fashion products from $10$ categories\footnote{We split the dataset into 55k, 5k, and 10k images as train, validation, and test sets (equally split over the $10$ categories). For testing, we generate 10k grid images with the test image set.}. 

\begin{figure}[t]
\begin{center}
\includegraphics[width=0.48\textwidth]{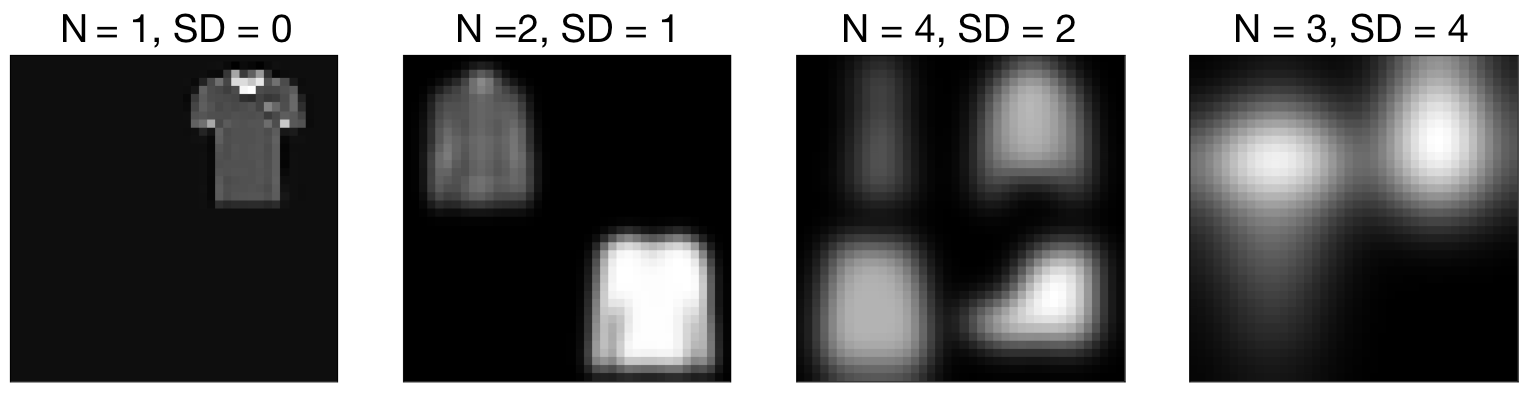}
\end{center}
\caption{Examples of stimuli used (with the number of objects, N, and the Gaussian blur standard deviations, SD).} 
\label{stim_eg}
\end{figure}

We want to assess the effects of cue-driven feature-based feedback on two object-feature manipulations. One, reducing the feature information through blurring. Two, introducing feature competition by adding more objects to the image. To do so, we construct $2\times 2$ grid ($40\,\text{px}\times 40\,\text{px}$) images, in which we can place $1$ to $4$ objects (category overlaps are allowed) and blur them with a Gaussian kernel with standard deviations varying uniformly from $0$ to $4$ pixels. Example images are shown in Figure~\ref{stim_eg}.

\subsection{Network Architecture}

The artificial neural network (ANN) accepts three inputs - the image, the cue, and the probe, and outputs whether the probe category is present in the image, as shown in Figure~\ref{model_describe}. We will now describe the sub-networks corresponding to the individual inputs, and their interactions.

\subsubsection{Nature of the object processing stream}

The object processing stream (OPS, in green in Figure~\ref{model_describe}) is a fully-connected ANN with one hidden layer. The input layer consists of $1600$ units, representing the $40\,$px x $40\,$px images. The output layer consists of $11$ units, $10$ of which give the probabilities of the categories present in the image. The $11\,$th unit is the out-of-sample detector which gives the probability that the input image does not belong to the space of images from the training set. This is done to prevent the OPS from make high-confidence errors on out-of-sample images. 

The hidden layer contains either $8$, $32$, or $3072$ rectified linear units (ReLU). An increase in the number of hidden layer units corresponds to an increase in neural capacity.

\subsubsection{Nature of the probe}

The probe is a one-hot encoding\footnote{\label{ft:2}Given $5$ categories, and $1,2,5$ being the categories of interest, an n-hot encoding (n$=3$ here) would be $[1,1,0,0,1]$} ($10$ units) of the category of interest. It is fed into another ANN with the output of the OPS (the category probabilities). This feedforward query ANN (fully-connected, in orange in Figure~\ref{model_describe}) has $200$ rectified linear units (ReLU) in its hidden layer. It has $2$ output units (Yes/No) which give the probability of the presence of the probe category in the image. 

\subsubsection{Nature of the cue}

\begin{figure}[t]
\begin{center}
\includegraphics[width=0.48\textwidth]{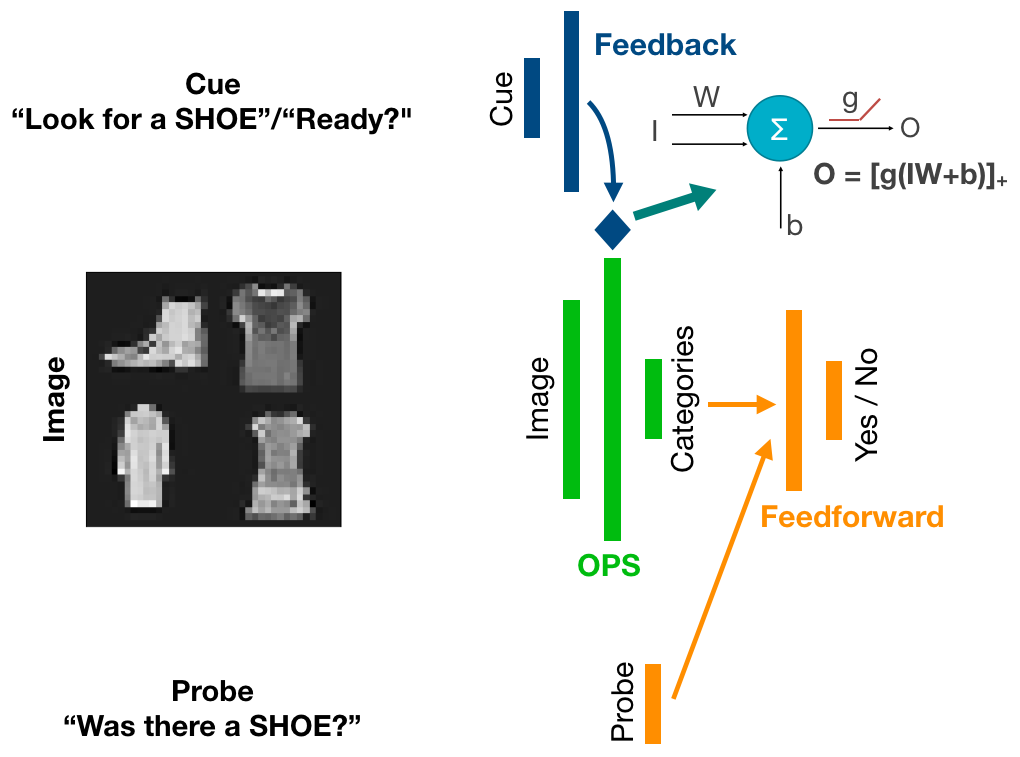}
\end{center}
\caption{The artificial neural network (ANN) designed to gauge the dependence of feedback-driven performance boosts due to the cue on the capacity limits of the object processing stream (OPS). The ANN outputs whether the probe category is present in the input image or not. The cue interacts with the OPS through bias (b) and/or gain (g) modulation of the hidden units of the OPS.} 
\label{model_describe}
\end{figure}

The cue consists of $11$ units, $10$ of which correspond to the `informative' cue as they correspond to the object categories. The $11\,$th unit corresponds to the `uninformative' cue (``Ready?'' as seen in Figure~\ref{model_describe}). The informative cue is the same as the probe. This cue, after being transformed into ‘feedback templates’, interacts with the OPS through bias and gain modulation, as explained next. This cue network is shown in blue in Figure~\ref{model_describe}.

\subsubsection{Cue--OPS interaction}
The responses $\mathbf{O}$ of the units in the hidden layer of the OPS are given by $\mathbf{O_{h}} = \mathbf{\left [ g_{h}(IW_{h}+b_{h}) \right ]_+}$, where $\mathbf{I}$ are the inputs, $\mathbf{W_{h}}$ are the input weights, and $\mathbf{b_{h}}$ and $\mathbf{g_{h}}$ are the biases and gains of the units. $[x]_+=x$ if $x> 0,$ \& $[x]_+=0$ if $ x \leq 0$. The feedback templates $\mathbf{b_c}$ and $\mathbf{g_c}$ are linear transformations of the cue, given by $\mathbf{b_c=z_cW_b} $ and $\mathbf{g_c=z_cW_g} $, where $\mathbf{z_c}$ is the one-hot encoding of the cue category. These templates are added to $\mathbf{b_{h}}$ and $\mathbf{g_{h}}$ respectively, causing either an additive or multiplicative boost in the units' responses. This interaction between the cue and the OPS was adapted from Lindsay \& Miller (2017).

\subsection{Network Training}

We now describe the input-output maps used in training the ANN shown in Figure~\ref{model_describe}. In each case, we learn the maps by minimising the cross-entropy between the network output and target probability distributions. We do so by using stochastic gradient descent (SGD) with Dropout regularisation~\cite{srivastava2014dropout}. The training is done in three steps.

\subsubsection{Training the OPS}

First, we train the parameters of the network marked in green in Figure~\ref{model_describe}. The inputs are the images mentioned in the Stimuli section. For each image, the target distribution, at the end of the OPS, is an n-hot encoding normalised to a unit vector, representing the $n$ unique object categories in the image. The $11\,$th unit of the OPS output is associated with random images\footnote{Uniformly random intensities are generated for all pixels. Then the image is scaled, blurred, and occluded to cover subspaces of interest better.}. To manipulate the representational capacity of the OPS, we train the OPS either with images containing a single object which is not blurred or with the full range of feature manipulations as shown in the Stimuli section. The representational capacity for the full range of feature manipulations is lower in the former case, which we refer to as low representational capacity here.

\subsubsection{Training the probe}

Second, we train the parameters of the network marked in orange in Figure~\ref{model_describe}. The OPS parameters are frozen. The inputs are the images with the full range of feature manipulations. The images are paired with correct or incorrect probe categories equally. The output of the ANN is a one-hot encoding of the correctness of the probe. 

\subsubsection{Training the cue}

Third, we train the parameters of the network marked in blue in Figure~\ref{model_describe}. All other ANN parameters are frozen. In the case of informative cues, where the cue category is the same as the probe category, the maps used in training the probe are paired with the respective cues. In the case of the uninformative cue, all the maps used in training the probe are paired with the cue. We train bias and gain modulation together, allowing for interactions between them.

\subsection{Evaluation metric}
Recognition performance is defined as the difference between the proportion of correct and incorrect assessment that the probe category exists in the input image. We assess the effects of imposing the two capacity limits on the recognition performance (True positives (TP) - False positives (FP)) boosts provided by the cues. If category information in the informative cue adds any functional (in terms of object categorisation) value, it should boost performance beyond the performance given by the uninformative cue (this boost is denoted by $\Delta$). 

\begin{figure}[t]
\begin{center}
\includegraphics[width=0.47\textwidth]{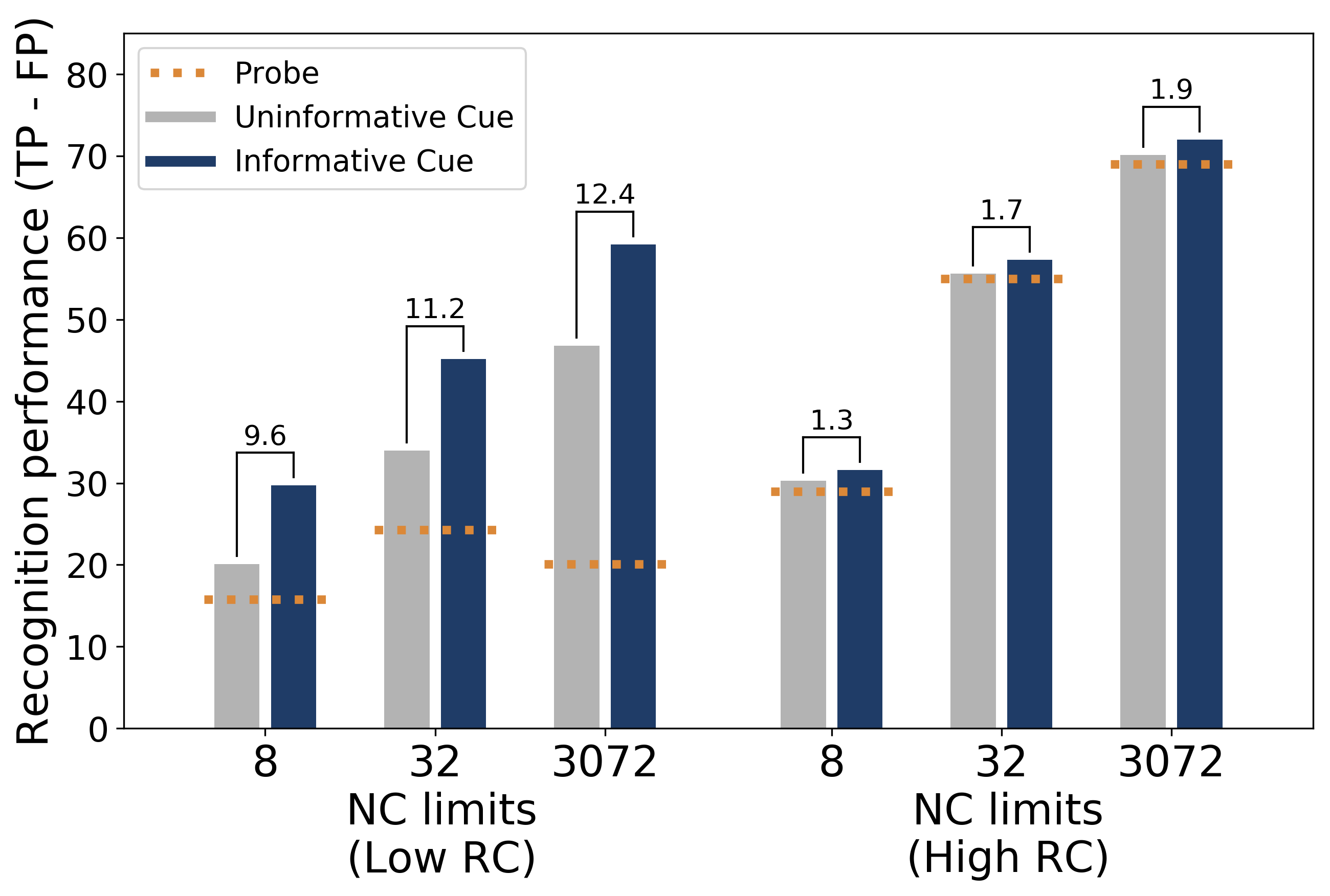}
\end{center}
\caption{Cue-driven recognition performance boosts as a function of the neural capacity (NC) and representational capacity (RC). The values of the boosts ($\Delta_{avg}$) given by the informative cue beyond the uninformative cue, are mentioned for NC/RC pair.  As the neural capacity is reduced, the informative cue provides lower performance boosts. Given a fixed neural capacity, the informative cue provides higher performance boosts when the representational capacity is reduced.} 
\label{results1}
\end{figure}

\section{Results and Discussion}

We evaluate the recognition performance on the full range of feature manipulations (number of objects and the strength of blurring), and in the case of joint training of the gain and bias modulation (in the cue ANN). The accuracies for recognising single objects in the image when the object processing stream (OPS) is trained on single objects in the image (number of hidden units mentioned in brackets) are $82.4\%\, (3072)$, $75.7\%\, (32)$, and $54.8\%\, (8)$. So, the representational capacity for single objects reduces with a reduction in neural capacity. 

The recognition performance of the probe-only, the uninformative cue, and the informative cue cases are mentioned in Figure~\ref{results1}. As seen in the figure, the informative cue provides higher recognition performance than the uninformative cue and the probe-only case when the representational capacity is reduced. The uninformative cue boosts performance over the probe-only case when the representational capacity is low. This performance boost could be a result of boosting the overall activity of the hidden units (through bias/gain) that provide reliable differences in activity for the object categories, in the case of the images with feature manipulations.

Trends observed in Figure~\ref{results1} are preserved if we vary only the number of objects (given $3072$ OPS hidden units, $\Delta_{avg,RC\uparrow} = 1.5\%$, $\Delta_{avg,RC\downarrow} = 11.7\%$; if $n_{obj}=4$ , $\Delta_{RC\uparrow} = 2.3\%$, $\Delta_{RC\downarrow} = 15.5\%$) or the strength of blurring in the test images (given $3072$ OPS hidden units, $\Delta_{avg,RC\uparrow} = 1.2\%$, $\Delta_{avg,RC\downarrow} = 5.4\%$; if blur SD $=4\,$px, $\Delta_{RC\uparrow} = 2.3\%$, $\Delta_{RC\downarrow} = 21.4\%$). 

So, cue-driven feature-based feedback seems to be useful for recognising objects subject to feature blurring and/or competition, only when the features required to classify those objects cannot be fully represented in the object processing stream. Intuitively, if the OPS can represent all the category-specific information about an object given the implicit representational limits imposed by the neural capacity, such feedback should not be able to add to the recognition performance.

\begin{figure}[t]
\begin{center}
\includegraphics[width=0.48\textwidth]{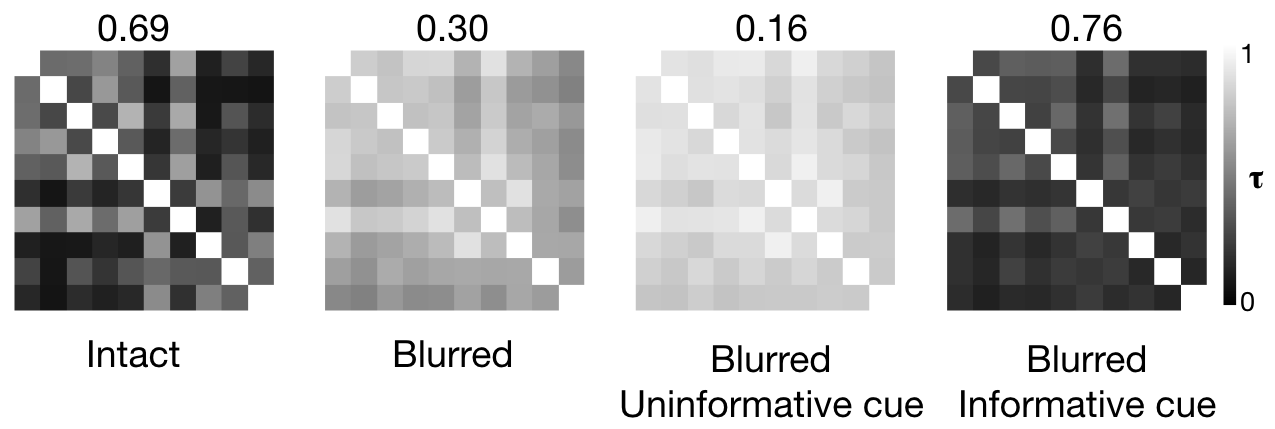}
\end{center}
\caption{The effect of cuing on the representation of the relevant object in the OPS hidden layer ($3072$ units and low RC). The standard deviation of blurring is $4\,$px. The category-level RSMs for the activities in the hidden layer for the mentioned cases are shown.  The categoricality indices for each RSM are shown. The informative cue makes the representations in the hidden layer selectively more distinct for the relevant object.} 
\label{blur-rsm-comp}
\end{figure}

\subsection{Influence of the trained feedback}

How does the external cue influence the representation of the relevant object in the hidden layer? 

To assess this influence, we generate category-level representational similarity matrices (RSM, based on Kendall's $\tau$ correlation) for the case of a single object in the grid, which is either presented intact, with blurring (with SD $=4\,$px), with blurring and the uninformative cue, or with blurring and the informative cue. We define the \textit{categoricality index} of the RSM as the difference between the mean values of the diagonal and off-diagonal elements. As seen in Figure~\ref{blur-rsm-comp}, the informative cue makes the representation of the relevant object more distinct, making it accessible to the output of the OPS. 

In Abdelhack \& Kamitani (2018), it was shown that neural representations of blurred objects in the human visual cortex are more similar to the corresponding intact object representations in a feedforward neural network than the blurred object representations in that neural network. They attributed this effect to top-down information interacting with the stimulus representations. This effect became stronger when a category cue was introduced. However, in our case, the neural representations of blurred objects (with either the informative or the uninformative cue in effect) are equally similar to the corresponding intact object representations (with no cue) and the blurred object representations (with no cue) ($\Delta \tau < 0.04$). This inconsistency will be probed in further work by using more complex and more biologically-plausible networks (such as convolutional neural networks and recurrent neural networks) for the object processing stream, which would make our network a better model of the human visual system.

\subsection{Comparison with tuning-based feedback}

A popular model to describe the effects of the cue on neuronal responses in the brain is the feature similarity gain model (FSGM)~\cite{martinez2004feature}. It claims that the neuronal response is multiplicatively scaled according to its preference to the properties of the attended (or task-relevant) stimuli. Such cue-driven `tuning-based' feedback was shown to boost object recognition performance (with multiple objects in a grid or overlaid) in Lindsay \& Miller (2017). We deployed tuning-based bias and gain modulation according to the mathematical framework outlined in Lindsay \& Miller (2017)\footnote{To implement the tuning-based modulations, the following steps are taken in Lindsay \& Miller (2017). Compute category-specific (averaged across multiple images) hidden layer activations. Mean- and variance-normalise the category values for each hidden unit to generate the feedback templates. Tune (multiplicative scaling only) these templates for bias (additive) or gain (multiplicative) modulation.}. We ran a grid search to compute the parameters to maximise the recognition performance boosts provided by tuning-based feedback over the probe-only case when representational capacity is low. Across the three neural capacities, the maximum performance boost observed was $3\%$. This is small compared to the boosts observed with the feedback trained with SGD as seen in Figure~\ref{results1}. This implies the trained feedback is not the same as tuning-based feedback.

Lindsay \& Miller did observe a higher performance using gradient-based feedback (of which feedback trained with SGD is the natural extension) than with tuning-based feedback. As also mentioned in their paper, this is not surprising as neuronal tuning is not necessarily a measure of neuronal function. It has been shown that category-selective responses of hidden units in ANNs do not imply that those units are relatively more important to the recognition of objects of those categories~\cite{morcos2018importance}. In fact, the greater the category-selectivity of the hidden units, the harder it is for the ANN to generalise to new data. 

\section{Conclusions}

In this work, we investigated the nature and usefulness of cue-driven feature-based feedback in recognising objects suffering from feature blurring and/or competition. We built an artificial neural network and asked how feature-based feedback can be deployed, and how its recognition performance boosts are dependent on neural and representational capacities of the object processing stream. We found that the feedback boosts performance only if the category-specific features about the objects cannot be fully represented in the base ANN. These representational limits are not dependent on the neural capacity but on the task demands on the object processing stream. The trained feedback does not resemble (but performs better than) tuning-based feedback which is based on the feature similarity gain model~\cite{martinez2004feature}.

To gauge the robustness of our observations, in subsequent work we will run these analyses on different datasets and architectures (such as convolutional neural networks). We shall also assess these effects for location-based feedback.

\section{Acknowledgements}
We thank Giacomo Aldegheri, Surya Gayet, and Nadine Dijkstra for their comments and suggestions.

This work was funded by the European Research Council (ERC) under the European Union's Horizon 2020 research and innovation program (Grant Agreement No. $725970$). This manuscript reflects only the authors' view, and the Agency is not responsible for any use that may be made of the information it contains.

\nocite{carrasco2011visual}
\nocite{bar2004visual}
\nocite{xiao2017fashion}
\nocite{lindsay2017understanding}
\nocite{srivastava2014dropout}
\nocite{morcos2018importance}
\nocite{abdelhack2018sharpening}

\bibliographystyle{apacite}

\setlength{\bibleftmargin}{.125in}
\setlength{\bibindent}{-\bibleftmargin}

\bibliography{thorat}

\end{document}